\documentstyle[12pt,aasms4]{article}

\def\gsim{\;\rlap{\lower 2.5pt
 \hbox{$\sim$}}\raise 1.5pt\hbox{$>$}\;}
\def\lsim{\;\rlap{\lower 2.5pt
   \hbox{$\sim$}}\raise 1.5pt\hbox{$<$}\;}

\def\ge{\;\rlap{\lower 2.5pt
 \hbox{$-$}}\raise 1.5pt\hbox{$>$}\;}
\def\le{\;\rlap{\lower 2.5pt
   \hbox{$-$}}\raise 1.5pt\hbox{$<$}\;}
\def\HI{\ion{H}{1}~}
\def\HII{\ion{H}{2}~}

\newcommand\beq{\begin{equation}}
\newcommand\eeq{\end{equation}}

\def\v{\vspace{-0.1in}}

\begin{document}

\Large \centerline{\bf Cosmic Hydrogen Was Significantly Neutral}
\centerline{\bf a Billion Years After the Big Bang}

\normalsize
\author{\bf J. Stuart B. Wyithe$^\dagger$ and Abraham Loeb$^\star$}
\medskip
\noindent
$\dagger$ School of Physics, The University of Melbourne, Parkville, Vic
3010, Australia \\ $\star$ Astronomy Dept., Harvard University, 60 Garden
Street, Cambridge, MA 02138, USA\\

\vskip 0.2in 
\hrule 
\vskip 0.2in 

{\bf The ionization fraction of cosmic hydrogen, left over from the big
bang, provides crucial fossil evidence for when the first stars and quasar
black holes formed in the infant universe.  Spectra of the two most distant
quasars known\cite{WB1} show nearly complete absorption of photons with
wavelengths shorter than the Ly$\alpha$ transition of neutral hydrogen,
indicating that hydrogen in the intergalactic medium (IGM) had not been
completely ionized at a redshift $z\sim6.3$, about a billion years after
the big bang.  Here we show that the radii of influence of ionizing
radiation from these quasars imply that the surrounding IGM had a neutral
hydrogen fraction of tens of percent prior to the quasar activity, much
higher than previous lower limits \cite{WB1}$^,$\cite{f2} of
$\sim0.1\%$. When combined with the recent inference of a large cumulative
optical depth to electron scattering after cosmological recombination from
the WMAP data\cite{KS1}, our result suggests the existence of a second peak
in the mean ionization history, potentially due to an early formation
episode of the first stars.}

The detection of a Ly$\alpha$ emitting galaxy at $z\sim6.56$ led to the
claim that the IGM must already be highly ionized at that
redshift\cite{Hu}. However, it was later shown\cite{H1} that the presence
of a small ionized region around the galaxy and broadening of the
Ly$\alpha$ line, would allow detection of the line feature in a fully
neutral IGM.  Luminous quasars offer an alternative probe of the IGM and
have higher luminosities, allowing acquisition of higher quality spectra.
The spectra\cite{WMJ1}$^,$\cite{WBC1} of SDSS~J1030+0524 at $z=6.28$ and
SDSS~J1148+5251 at $z=6.41$ show transmitted flux down to wavelengths
corresponding to the Ly$\alpha$ resonance of hydrogen at $z\sim 6.20$ and
$z\sim 6.32$ respectively, even though no flux is detected at somewhat
greater distances from the quasars\cite{WBC1}.  These redshift
displacements correspond to velocity shifts of order $\sim 3300~{\rm
km~s^{-1}}$, and imply ionized regions surrounding the quasars with
observed physical radii of $R_{\rm obs}=4.5$~Mpc and $4.7$~Mpc,
respectively.  The inference is robust since high ionization broad-lines
typically have velocity offsets from the true quasar redshift\cite{RVB1} of
only $\la 1000~{\rm km~s^{-1}}$.

We model the evolution of a fully ionized region (the so-called {\it
Str\"omgren sphere}) around a quasar that emits UV photons into a partially
neutral IGM.  Prior to the overlap phase of ionized hydrogen (\HII) regions
that signals the end of the reionization epoch\cite{BL1}, galaxies generated their
own small ionized regions filling a fraction of the IGM volume around the
quasar. We define the neutral
fraction, $x_{\rm HI}$, to be the mean fraction of hydrogen atoms 
(in a representative volume of the low-density IGM) 
that remain neutral prior to the quasar activity, 
and assume $x_{\rm HI}$ outside the Str\"omgren
sphere to remain constant during its expansion.  Since the luminosity
of a bright quasar is much larger than the stellar luminosity of
neighboring galaxies, a fully ionized quasar bubble expands into a
partially ionized IGM composed of isolated \HII regions. 

Even for the deepest available spectra, with lower limits on the Ly$\alpha$
({\it Gunn-Peterson}\cite{GP}) optical depth of $\tau_{\rm GP}\ga26$, the
IGM need only have a neutral fraction of $\sim 10^{-3}$ in order to produce
the observed absorption blueward of Ly$\alpha$\cite{WB1}.  However, the
value of $R_{\rm obs}\sim4.5$Mpc may be used to constrain larger values of
the neutral fraction in the range $0.001\leq x_{\rm HI}\leq1.0$. In the
early expansion phase, the dependence of the physical radius of the
Str\"omgren sphere, $R_{\rm p}$, on the quasar age\cite{WB1}$^,$\cite{Pe}, 
$t_{\rm age}$, may be crudely approximated as $R_{\rm p}\sim 7x_{\rm
HI}^{-1/3}\left[t_{\rm age}/10^7{\rm yr}\right]^{1/3}$Mpc, given the
production rate of ionizing photons that is characteristic of
SDSS~J1148+5251 and SDSS~J1030+0524. Clearly, the Str\"omgren sphere grows
larger when embedded in an IGM with a small neutral fraction.  
Defining $R_{\rm max}$ to be the maximum radius achieved in a fully neutral
IGM by the end of the quasar lifetime, $t_{\rm lt}$, the resulting a~priori
probability distribution is, $\left.dP/dR_{\rm p}\right|_{x_{\rm
HI}}=3x_{\rm HI}R_{\rm p}^2/R_{\rm max}^3$ for $R_{\rm p}<R_{\rm max}x_{\rm
HI}^{-1/3}$. In addition, the neutral fraction must be smaller than $x_{\rm
max}={\rm min}\left(\left[t_{\rm lt}/10^7{\rm years}\right]\left[R_{\rm
obs}/7{\rm Mpc}\right]^{-3},1\right)$ to allow the growth of the sphere to
$R_{\rm obs}$ within $t_{\rm lt}$.  Using a flat logarithmic prior for
$x_{\rm HI}$, we find the cumulative a~posteriori probability distribution,
$P(<x_{\rm HI})={\rm min}(x_{\rm HI},x_{\rm max})/x_{\rm max}$. Since
estimates of quasar lifetimes\cite{MA1} are bracketed in the range
$10^6-10^8$ years, we get $x_{\rm max}\sim1$ and conclude that $x_{\rm
HI}\ga0.1$ ($0.01$) with 90\% (99\%) confidence.

To better quantify this simple argument, we write the general equation for
the relativistic expansion of the {\em co-moving} radius [$R=(1+z)R_{\rm
p}$] of the quasar \HII region\cite{WB1} in an IGM with a neutral filling
fraction $x_{\rm HI}$ (fixed by other ionizing sources),
\begin{equation}
\label{Vev2}
\frac{dR}{dt}=c(1+z)\left[\frac{F_\gamma\dot{N}_{\rm
ion} - \alpha_{\rm B}C F_{\rm m} x_{\rm HI}\left(\bar{n}_0^{\rm
H}\right)^2 \left(1+z\right)^3 \left({4\pi\over 3}R^3\right)}
{F_\gamma\dot{N}_{\rm ion} + 4\pi R^2 \left(1+z\right) c F_{\rm m} x_{\rm
HI}\bar{n}_0^{\rm H}}\right],
\end{equation}
where $c$ is the speed of light, $\bar{n}_0^{\rm H}$ is the mean number
density of protons at $z=0$, $\alpha_{\rm
B}=2.6\times10^{-13}$cm$^3$s$^{-1}$ is the case-B recombination coefficient
at the characteristic temperature of $10^4$K, and $\dot{N}_{\rm ion}$ is
the rate of ionizing photons crossing a shell at the radius of the \HII region at time
$t$.  We use the distribution derived from numerical simulations for the
over-densities in gas clumps\cite{MHR1}, and calculate the mean free path
$d(\Delta_{\rm c})$ for ionizing photons\cite{MHR1}$^,$\cite{BL2}.
Following Barkana \& Loeb\cite{BL2}, we then find the critical overdensity
$(\Delta_{\rm c})$ at which a fraction $F_\gamma \equiv 0.5 =
\exp\left[-R_{\rm p}/d(\Delta_{\rm c})\right]$, of photons emitted do not
encounter an overdensity larger than $\Delta_{\rm c}$ within the \HII
region.  We also compute the mass fraction $F_{\rm m}$ ($\sim1$) of gas
within $R_{\rm p}$ that is at over-densities lower than $\Delta_{\rm c}$.
Finally, we calculate the clumping factor in the ionized regions,
$C(R)\equiv\langle\Delta^2\rangle/\langle\Delta\rangle^2$, where the
angular brackets denote an average over all regions with
$\Delta<\Delta_{\rm c}$. For $x_{\rm HI}=1$ and $dR_{\rm p}/dt\ll c$,
equation~(\ref{Vev2}) reduces to its well-known
form\cite{Pe}$^,$\cite{SG1}$^{-}$\cite{MR1}.  The expansion of the quasar
Str\"omgren sphere would change in an overdense region of the IGM.  However
the density contrast due to infall has been shown\cite{BL2} to be small
($\sim 1\%$) on scales of several Mpc around the massive hosts of the
bright SDSS quasars at $z\sim6$. Numerical experiments showed our results
to be insensitive to infall.  The emission rate of ionizing photons
$\dot{N}_{\rm ion}$ in equation~(\ref{Vev2}) is computed at $t'=t-t_{\rm
delay}$ to account for the finite light travel time between the source and
the ionization front. The delay $t_{\rm delay}$ is derived from the
relation $R = \int_{t_R-t_{\rm delay}}^{t_R} cdt'\left[1+z(t')\right]$,
where $t_R$ is the time when the photon crosses the co-moving radius
$R$. The use of the Telfer et al.\cite{T1} spectral template implies
$\dot{N}_{\rm ion}=6.5\times10^{57}$s$^{-1}$ for SDSS J1030+0524 and
$\dot{N}_{\rm ion}=10.0\times10^{57}$s$^{-1}$ for SDSS J1148+5251, while
the template of Elvis et al.\cite{E1} implies lower values of $\dot{N}_{\rm
ion}\sim1.3\times10^{57}$s$^{-1}$ and $\dot{N}_{\rm
ion}\sim2.0\times10^{57}$s$^{-1}$, respectively\cite{WB1}.  

Our model for the evolution of the Str\"omgren sphere includes the quasar
formation history.  We compute the time dependent ionizing flux governed by
black-hole growth through accretion and mergers.
Following the observational inference that the relation between bulge
velocity dispersion and black hole mass $M_{\rm bh}$ does not evolve with
redshift\cite{SGS1}, we extrapolate the present-day $M_{\rm bh}$--$M_{\rm
halo}$ relation\cite{fe} using the dependence of virial velocity on
redshift\cite{BL1} to obtain $M_{\rm halo} = 1.5\times 10^{12}M_\odot
\left({M_{\rm bh}}/{10^9 M_\odot}\right)^{3/5}
\left[(1+z)/7\right]^{-3/2}$. The luminosities of the SDSS quasars imply
black-hole masses of $\sim2\times10^9$M$_\odot$ accreting at their
Eddington rate\cite{WMJ1}, leading to inferred host dark-matter halos of
$M_{\rm halo}\sim2\times10^{12}$M$_\odot$.  The inference of such a massive
halo at this early epoch is supported by the signature of gas infall in the
spectra of high redshift quasars\cite{BL2}, and the inference of a large
molecular mass and velocity width in the host galaxy of
SDSS~J1148+5251\cite{WBC1}.

We begin with two dark-matter halos of comparable mass $M_1$ and $M_2$ that
merge to form the host dark-matter halo with a mass $M_{\rm
halo}=(M_1+M_2)$ at a redshift corresponding to one quasar lifetime prior
the observed redshift, and construct merger trees for each of the
progenitor halos using the algorithm described by Volonteri et
al.\cite{VHM1}. We use the $M_{\rm bh}$--$M_{\rm halo}$ relation to find
the mass of the black-hole at the center of each dark-matter halo in the
merger tree. When there is a major merger (having a progenitor mass ratio
$<2$), we assume that the black-holes merge and that the accretion of the
mass deficit relative to the $M_{\rm bh}$--$M_{\rm halo}$ relation produces
the limiting Eddington luminosity with a radiative efficiency of
$\epsilon=10\%$.  The resulting quasar lifetime is $t_{\rm
lt}=4\times10^7({\epsilon}/{0.1})\ln{\left[{M_{\rm
halo}^{5/3}}/({M_1^{5/3}+M_2^{5/3}})\right]}\mbox{ years}$.  The ionizing
photon emission rate $\dot{N}(t)$ is taken to have an exponential time
dependence, $\dot{N}_{\rm ion}(t)=\dot{N}_{0} e^{-t/t_{\rm lt}}$, with
$\dot{N}_0 \propto M_{\rm bh}$.  We assume that all quasar episodes in the
merger tree contribute to the volume of the observed Str\"omgren sphere,
which is centered on the most massive halo in the tree.  A similar
prescription for quasar activity was shown to be consistent with the
observed number counts of high redshift quasars\cite{VHM1}.  For a major
merger resulting in the observed quasar activity, the lifetime given by
this approach is $\sim 10^7$ years.  This compares favorably with a variety
observational estimates of quasar lifetimes\cite{YT1}$^-$\cite{MS1}.
To cover the
full range of uncertainty we consider modifications to our fiducial model
in which the quasar lifetime is multiplied by a factor $f_{\rm lt}$ between
0.1--10, resulting in lifetimes for the observed quasars of $f_{\rm
lt}t_{\rm lt}=10^6$--$10^8$ years.

We have produced 300 realizations of the merger tree, and computed the
evolution of the Str\"omgren sphere in each case for the full range of
neutral fractions $0.001\le x_{\rm HI}\le1.0$, allowed by the Gunn-Peterson
optical depth\cite{WB1}.  We find that quasar activity associated with the
hierarchical build up of the host galaxy produces an \HII region with a
typical size of $\sim2x_{\rm HI}^{-1/3}$Mpc, comparable to the observed 
radii if $x_{\rm HI}$ is of order unity.  
Figure~\ref{fig1} shows the
conditional a~priori probability distributions of $R_{\rm p}$ for quasars
like SDSS J1030+0524, assuming $x_{\rm HI}=1$ and two different values of
$f_{\rm lt}$. The observed radii are consistent with lifetimes close to the
fiducial case of $f_{\rm lt}=1$. The plotted distribution of $R_{\rm p}$
implies that the hierarchical evolution of early quasars leads to a dearth
of small Str\"omgren sphere radii for the {\em observed} SDSS quasars. This
is in contrast to the simplified monolithic model of quasar
formation\cite{CH1}$^,$\cite{WB1}, for which the \HII region has zero volume 
at the time when the quasar turns on [$R(t_{\rm age}=0)=0$] and the
distribution extends down to $R_{\rm p}=0$. 

To constrain the neutral fraction, we have computed the likelihood
of observing $R_{\rm obs}=4.5$Mpc around SDSS~J1030+0524 and $R_{\rm
obs}=4.7$Mpc around SDSS~J1148+5251 as a function of $x_{\rm HI}$ and
$f_{\rm lt}$. The results are plotted
in Figure~\ref{fig2}. The upper panels show the locus of most likely values
(thick line) as well as likelihood contours at 0.1 of the peak value
(dashed lines). 
The fiducial lifetime ($f_{\rm lt}$) favors $x_{\rm HI}\sim1$, while with $f_{\rm
lt}\ll1$ the distributions for $R_{\rm p}$ lie substantially below the
observed value of $4.5$Mpc, making smaller values of $x_{\rm HI}$ more
likely.  Extrapolation of the most likely contour for the Elvis et
al.\cite{E1} template yields $f_{\rm lt}\sim 2 x_{\rm HI}$.  This implies
that $x_{\rm HI}\sim 10^{-3}$ would require a lifetime as short as
$2\times10^4$ years, which is ruled out by variability properties of
quasars in SDSS\cite{MS1}.  The a~posteriori probability distributions 
for $x_{\rm HI}$ given $f_{\rm lt}$ are plotted in
the lower panels of Figure~\ref{fig2} and robustly yield the constraint $x_{\rm
HI}\gsim0.01$. For $f_{\rm lt}\ga0.3$, we find $x_{\rm HI}\gsim0.1$ and
$x_{\rm HI}\gsim0.4$ at the 90\% level for the Elvis et al.\cite{E1} and
Telfer et al.\cite{T1} spectra, respectively. The fiducial model with
$f_{\rm lt}=1$ yields corresponding constraints of $x_{\rm HI}\gsim0.3$ and
$x_{\rm HI}\gsim0.6$.  

The inference of a large neutral fraction at $z\sim6.3$ presents a
challenge to theories of cosmological reionization when combined with the
large optical depth\cite{KS1} to electron scattering after cosmological
recombination, $\tau_{\rm es}=0.17\pm0.04$.  Consider a toy model in which
the universe was partially reionized at a high redshift $z_{\rm reion,I}$,
leaving a neutral fraction $x_{\rm HI}$ until complete reionization was
reached at $z_{\rm reion,II}\sim6.25$.  The optical depth is then
$\tau_{\rm es}=0.04 + 0.002(1-x_{\rm HI})\left[(1+z_{\rm
reion,I})^{3/2}-19.5\right]$.  If the IGM ionization fraction increased
monotonically, then given $\tau_{\rm es}>0.13$, the universe needed to be
reionized earlier than $z_{\rm reion,I}\sim18$ or $z_{\rm reion,I}\sim24$
assuming the Elvis et al.\cite{E1} and Telfer et al.\cite{T1} spectra,
respectively.  If $\tau_{\rm es}=0.17$ and $x_{\rm HI}> 0.7$, as implied by
the Telfer et al.\cite{T1} spectrum, then a monotonic reionization history
requires significant reionization at an implausibly high redshift
($z\ge30$). In this case, a more plausible alternative is a non-monotonic
history with an early reionization peak, possibly due to the formation of
massive population-III stars\cite{WL2}$^,$\cite{C1}.

\noindent {\it Correspondence and requests for materials to Abraham Loeb.}

\vskip 0.2in
\noindent
{ACKNOWLEDGEMENTS.} This work was supported in part by grants from ARC,
NSF and NASA.

\vskip 1in

\begin{figure*}[htbp]
\caption{\label{fig1} Predicted probability for observing different radii
of the ionized region around the quasar SDSS J1030+0524.  The differential
a~priori probability distribution is plotted for $x_{\rm HI}=1$ and two
quasar lifetimes: $f_{\rm lt}=0.1$ ({\em dashed}) and $f_{\rm lt}=1.0$
({\em solid}). The {\em light} and {\em dark} curves show results
calculated based on the Elvis et al.\cite{E1} and Telfer et al.\cite{T1}
template spectra, respectively. The distributions may also be used to
approximate the behavior with $x_{\rm HI}<1$ by replacing $R_{\rm p}$ with
$R_{\rm p}x_{\rm HI}^{-1/3}$.  Values of $R_{\rm p}$ significantly larger
than observed still allow transmission of Ly$\alpha$ flux at a detectable
level because: (i) the optical depth due to the damping wing of the
IGM\cite{ME1} is only important near the boundary of the \HII region, and
(ii) the optical depth at $R_{\rm p}$ due to resonant absorption within the
\HII region\cite{BL2} is in the range $\tau_{\rm reson}=(1-8)\times(R_{\rm
p}/4.5~{\rm Mpc})^2$.  A dense \HI cloud near $R_{\rm obs}$ could produce a
large damping wing and lead to a systematic underestimation of
$R_p$. However, such a cloud would need a column density
$\ga10^{21}\,$cm$^{-2}$ ($\ga10^{22}\,$cm$^{-2}$) to produce a Ly$\alpha$
optical depth $>26$ over 10\% (30\%) of the spectral range covered by the
\HII region. This requires a damped Ly$\alpha$ absorber, whose existence is
highly improbable within the narrow redshift interval under consideration,
$\Delta z\sim0.1$.  Our calculation assumes that the ionization front is
thin. The spectrum-averaged mean-free path for the ionizing quasar photons
is $\lambda\sim 1.5x_{\rm HI}^{-1}(1+z)^{-3}$Mpc, which may be compared to
the bubble radius $R_{\rm p}\sim4.5x_{\rm HI}^{-1/3}$Mpc, to yield the
fractional thickness of the ionization front $({\lambda}/{R_{\rm p}})\sim
10^{-3}\left({R_{\rm p}}/{4.5{\rm Mpc}}\right)^{-1}x_{\rm
HI}^{-2/3}\left[{(1+z)}/{7.3}\right]^{-3}$.  We also note that the ionized
region need not be spherical; if the quasar radiation is beamed, its
luminosity per unit solid-angle along the observer's line-of-sight is still
measured, and equation~(\ref{Vev2}) therefore describes the extent of the
ionized region along the line-of-sight. Throughout our calculations, we
have adopted the best-fit cosmological parameters derived from the WMAP
data\cite{SB1}.  }
\end{figure*}

\clearpage

\begin{figure*}[htbp]
\epsscale{.9}  
\plotone{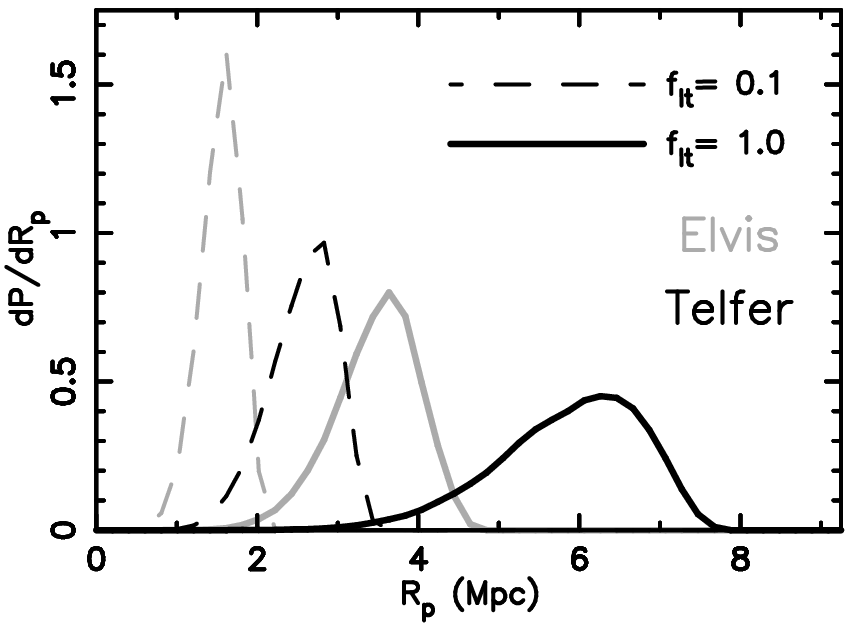}
\end{figure*}

\begin{figure*}[htbp]
\caption{\label{fig2} Likelihood for the inferred neutral fraction of the
IGM, assuming different quasar lifetimes.  We show contours of likelihood,
${\rm L}$, for $x_{\rm HI}$ and $f_{\rm lt}$ ({\em top}) and a~posteriori
cumulative probability distributions for $x_{\rm HI}$ ({\em
bottom}). Cumulative distributions are shown for two different quasar
lifetimes: $f_{\rm lt}=0.1$ ({\em dashed}) and $f_{\rm lt}=1.0$ ({\em
solid}). The {\em light} and {\em dark} lines show results where the quasar
ionizing photon rate was specified by the Elvis et al.\cite{E1} and Telfer
et al.\cite{T1} spectra, respectively.  The a~posteriori probability is
$\left.\frac{dP}{dx_{\rm HI}}\right|_{R_{\rm obs}} \propto
\left[\frac{dP_{1030}}{dR_{\rm p}}\left(\left.R_{\rm p}=4.5{\rm
Mpc}\right|x_{\rm HI}\right)\times\frac{dP_{1148}}{dR_{\rm
p}}\left(\left.R_{\rm p}=4.7{\rm Mpc}\right|x_{\rm
HI}\right)\right]\frac{dP_{\rm prior}}{dx_{\rm HI}}$, where $\frac{dP_{\rm
prior}}{dx_{\rm HI}}$ is the prior probability for $x_{\rm HI}$, assumed to
be flat in logarithmic bins within the range $0.001\le x_{\rm HI}\le1.0$,
and $\frac{dP_{1030}}{dR_{\rm p}}$ and $\frac{dP_{1148}}{dR_{\rm p}}$ are
the a~priori probability distributions for $R_{\rm p}$ in quasars like
SDSS~J1030+0524 and SDSS~J1148+5251, respectively. A flat logarithmic prior
is the natural choice for $\frac{dP_{\rm prior}}{dx_{\rm HI}}$, because
$x_{\rm HI}$ may be thought of as a ratio of independent quantities (the
number of ionizing photons and the number of baryons) that would themselves
have linearly distributed prior probabilities. The alternative use of a
linear prior in $x_{\rm HI}$ leads to more stringent limits than those
presented here.}
\end{figure*}

\begin{figure*}[htbp]
\epsscale{.9}  
\plotone{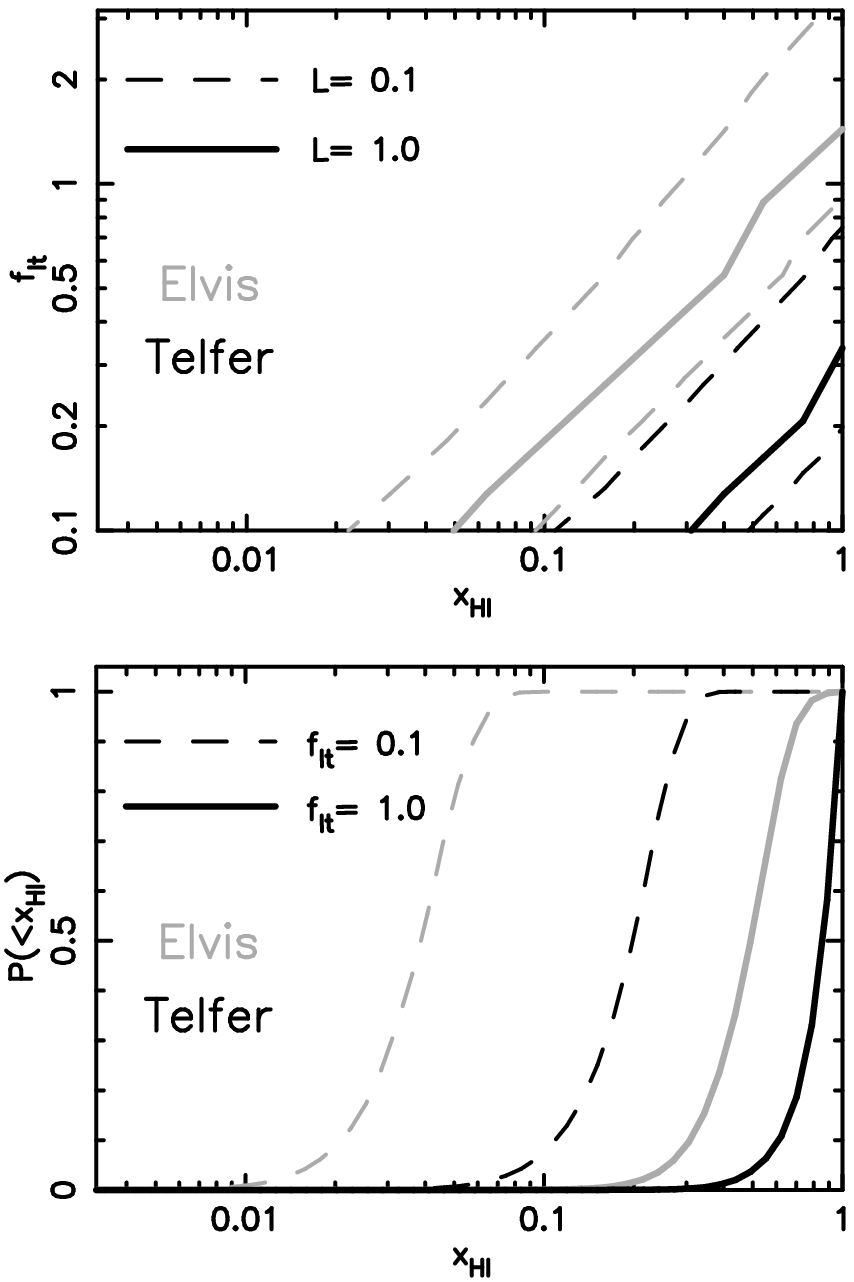}
\end{figure*}

\end{document}